\documentclass[12pt]{article}
\usepackage[draft]{hyperref}
\usepackage{amsmath}
\usepackage{graphicx}
\usepackage{authblk}

\begin{document}


\title{Dispersion Compensation using a Prism-pair}


\author{Yaakov Shaked, Shai Yefet and Avi Pe'er$^{*}$}
\affil{Department of physics and BINA Center of nano-technology, Bar-Ilan university, Ramat-Gan 52900, Israel
	
	$^*$e-mail: avi.peer@biu.ac.il}

\date{Dated: \today}

\maketitle

\begin{abstract}
     A simple and intuitive formulation is reviewed for the Brewster prism-pair - A most common component in spectroscopy-oriented experiments using ultrashort pulses. This review aims to provide students and beginners in the field of spectroscopy with a unified description of a major experimental component. The total spectral phase experienced by a broadband light field is calculated after passing through a pair of Brewster-cut prisms, demonstrating the flexibility of the prism pair to provide tuned, low-loss control of the dispersion and spectral phase experienced by ultrashort pulses.
\end{abstract}


\section{Introduction}

The generation of ultrashort pulses \cite{firstHeNemodelocking,firstKLM} revolutionized the field of molecular spectroscopy \cite{femtochemistry}. Not only that ultrashort pulses enable to observe molecular motion in time \cite{tannor1986coherent, shapiro1996coherent, hansch2006nobel}, but they allow also to manipulate their dynamics \cite{cr00025a006,Assion30101998,aharonovich2016coherent} by controlling the spectral properties of the pulses. For this end, many types of pulse-shaping techniques and configurations are very common in spectroscopy-oriented experiments. When complete control of the spectral amplitude and phase is required, a general Fourier-domain pulse-shaper \cite{Weiner:88} can be used, which provides independent control of phase and amplitude for each frequency component of the light (e.g. with a spatial light modulator \cite{dayan2005nonlinearSLM, weiner2011ultrafastSLM} or deformable mirrors \cite{Weinacht2001333}). However, most applications require only much simpler control of the group delay dispersion (GDD) and higher order dispersion to compensate for the dispersive effect experienced by an ultrashort pulse when passing through optical media and setups. For these applications, the high internal loss and technical complexity of a general pulse-shaper are a burden, and simpler configurations are commonly used, such as the grating-pair \cite{oron}, chirped mirrors \cite{vered2012twoGDDmirror} or the Brewster prism-pair \cite{Gershgoren:03} - the subject of this mini-review. Prism-pairs can provide tuned compensation with ultra low-loss of up to two orders of dispersion (GDD and sometimes an additional higher order), along with simple amplitude-control using a slit or transmission mask in the dispersive arm. Due to the low loss of the prism-pair, it is often the main 'tool of choice' for intra-cavity applications \cite{PrismPairCavity2007Theory, fork1987compressionPrismCavity, spence1991femtosecond}, low light level spectroscopy \cite{ogilvie2006useLiveCoherent, entenberg2011setupNature, kobayashi1990relaxation}, and quantum optics experiments \cite{shaked2015octave, shaked2014observing, BroadbandBiPhotonSesarn2010Harris, Peer2006designPrism, 1367-2630-16-3-033017}. This mini-review presents a simple and intuitive formulation of the total wavelength-dependent phase accumulated by light passing through a prism-pair, and demonstrates how this major component is used for tuned, precise control of dispersion and spectral phase.

\section{A single prism} \label{singleprism}

\begin{figure}
\centerline{\includegraphics[width=8cm]{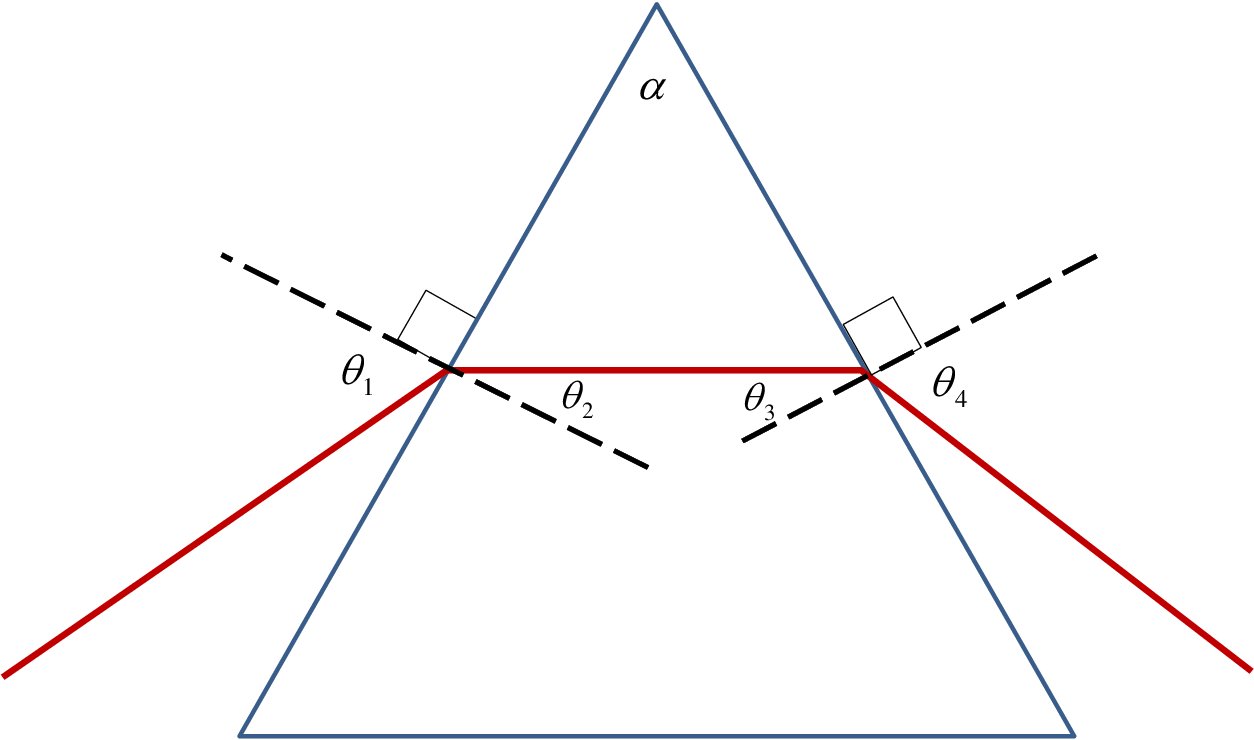}}
\caption{\label{minimumdeviation} Symmetry relations between the refraction angles in a prism. At minimum deviation, the entrance and exit angles are equal and the ray propagates through the prism parallel to its base.}
\end{figure}

 The analysis starts by reviewing the geometry of a single prism. When a ray passes through a prism at minimum deviation, the angles of refraction through the prism are symmetric \cite{HechtEugeneOptics}, as illustrated in Fig. \ref{minimumdeviation}, resulting in:
\begin{equation}
\begin{aligned}
\theta_{1}&=\theta_{4}\\
\theta_{2}&=\theta_{3}=\frac{\alpha}{2}  \label{symetric},
\end{aligned}
\end{equation}
where $\alpha$ is the apex angle of the prism. If the entrance angle matches with the Brewster angle $\theta_{B}$ (for a certain wavelength $\lambda_{0}$), the refraction angles obey
\begin{equation}
\begin{aligned}
\theta_{B}=\theta_{1}=\theta_{4}\\
\theta_{B}+\theta_{2}=\frac{\pi}{2},  \label{brewster}
\end{aligned}
\end{equation}
and the relation between the apex angle $\alpha$ and Brewster angle can be expressed as
\begin{equation}
\begin{aligned}
\alpha=2\theta_{2}=2\left(\frac{\pi}{2}-\theta_{B}\right)=\pi-2\theta_{B}\qquad  \label{apex}
\end{aligned}
\end{equation}

\section{The optical path through a prism-pair} \label{opticalpath}

\begin{figure}
\centerline{\includegraphics[width=13cm]{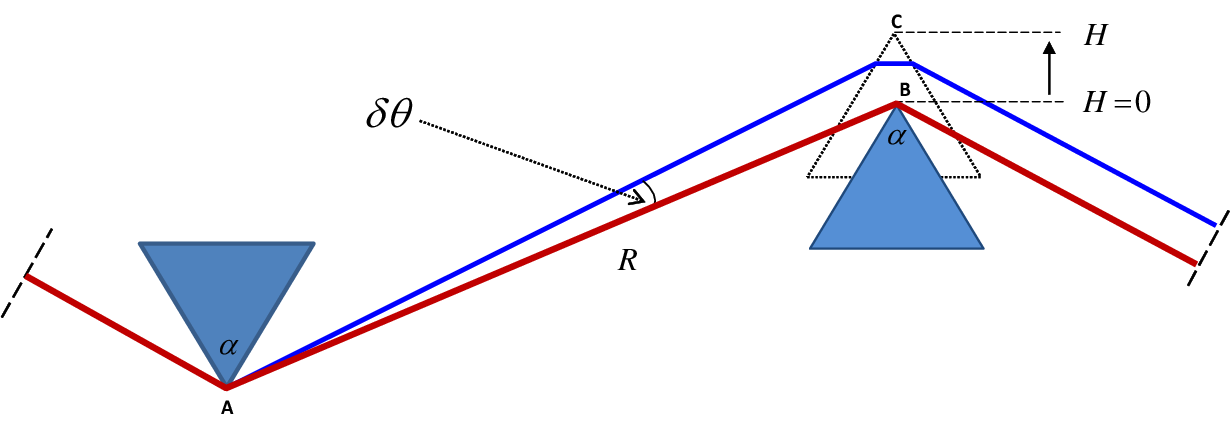}}
\caption{\label{coordinates} Geometry of a prism-pair, defined using two degrees of freedom: the separation between the prisms R and the prism insertion H.}
\end{figure}

Our aim now is to calculate the total optical path (and phase) through a prism-pair, which will allow to derive the dispersion experienced by the passing optical pulse. The coordinate system used for the prism-pair has two degrees of freedom (see Fig. \ref{coordinates}). R is the separation between the prisms (segment AB), and H is the penetration of the prism into the beam (segment BC). The red line represents a Brewster ray (at the design wavelength $\lambda_{0}$) that enters and exits both prisms at Brewster angle $\theta_{B}$. The blue line is the deviated ray at $\lambda\neq\lambda_{0}$, which deviates from the Brewster ray by an angle $\delta\theta$ due to the prism dispersion.\\

\subsection{Wavefront calculation} \label{wavefront}

\begin{figure}
	\centerline{\includegraphics[width=13cm]{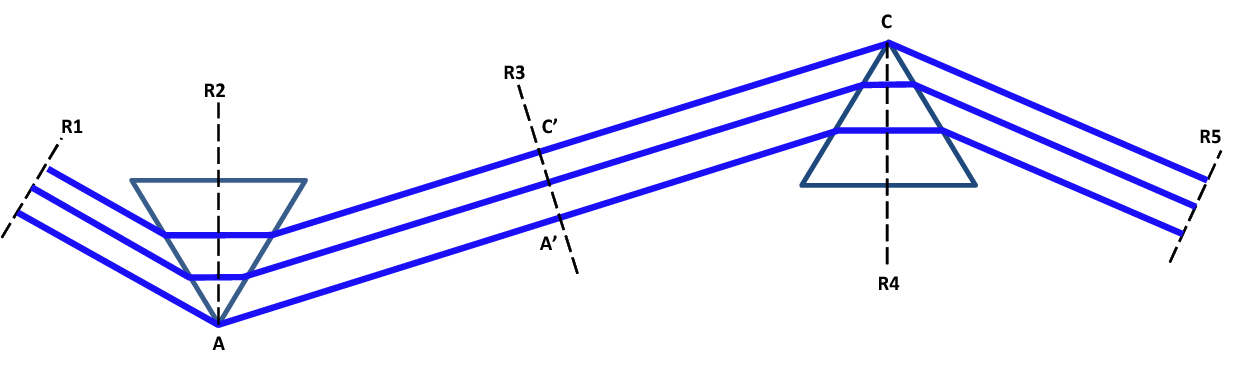}}
	\caption{\label{wavefronts} Different wavefronts (black dashed lines) of equal phase in a plane-wave beam passing through a prism pair.}
\end{figure}

To calculate the phase accumulated by an arbitrary wavelength $\lambda$ (blue line) through the prism-pair, one usually calculates the optical path of the deviated beam along a \emph{continuous} ray, starting from point A, through multiple refractions in the 2nd prism until a final reference plane perpendicular to the beam, located after the 2nd prism where all the colors are parallel \cite{Yang2001381,prismpairEXACT}. This method results in somewhat complicated expressions for the optical path, leaving little room for intuition. An elegant alternative (outlined in Fig. \ref{wavefronts}) that avoids calculations of multiple refractions is to use the concept of wave-fronts for a \emph{non-continuous} ray. This concept was originally presented in \cite{prismpairapprox} and calculated to include only the prisms separation R. The following generalizes the calculation to include also the prism insertion H.\\

\begin{figure}
\centerline{\includegraphics[width=13cm]{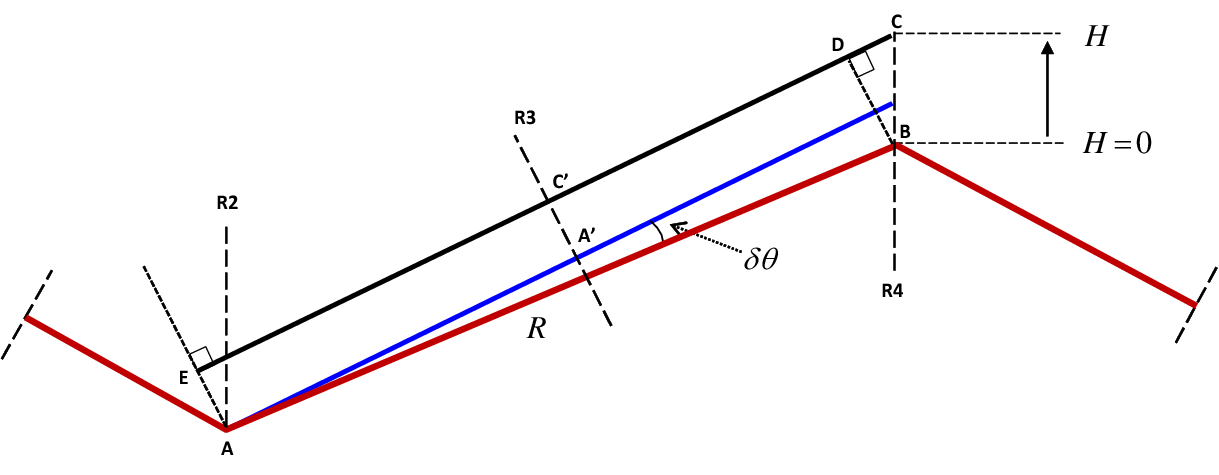}}
\caption{\label{planes} Complete optical path through a prism-pair. Following a non-continuous ray by using the concept of wavefronts, the optical path through the entire system can be reduced to the segment EC.}
\end{figure}

Within a plane-wave beam, Each ray has the same optical path (and phase) through the prism-pair (see Fig. \ref{wavefronts}), starting from reference plane R1 to R5 (all the rays have the same wavelength $\lambda$). Note that the optical phase is constant across each of the reference planes (dashed black lines), since each of them coincides with the wavefront perpendicular to the beam. The above argument is true not only for the Brewster ray $\lambda_{0}$, but for any other wavelength $\lambda$. Since the wavefront R1 is common for all wavelengths, shifting its position only results in a simple optical delay, hence, the position of R1 can be chosen freely to pass through the point A. The same argument applies also to R5, which can be chosen to pass through point C. Consequently, for calculating the geometric dispersion, it is sufficient if the optical path through the entire prism-pair is reduced to the segment:
\begin{equation}
\begin{aligned}
P=AA^{'}+CC^{'}  \label{aacc}.
\end{aligned}
\end{equation}
Note that although this path is composed of free-space rays without refractions, it already takes into account refraction of the planar beam in the material. As illustrated in Fig. \ref{planes} (which shows the rays without the prisms for clarity), the optical path in Eq. \ref{aacc} is identical to the segment $EC$, where the point C is the vertex of the 2nd prism. The conclusion is that the optical path of the deviated ray through the entire prism pair is simply given by
\begin{equation}
\begin{aligned}
P=EC=ED+DC&,  \label{EC}
\end{aligned}
\end{equation}
where $ED=R\cos\delta\theta$, and $DC=H\cos\theta_{H}$ ($\theta_{H}=\angle DCB$).\\

\begin{figure}
\centerline{\includegraphics[width=8cm]{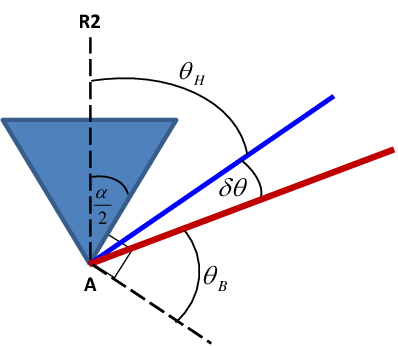}}
\caption{\label{thetaH} Definition of the exit angles in the 1st prism.}
\end{figure}

Since $EC$ is parallel to $AA^{'}$, and R2 is parallel to R4, one can see that $\theta_{H}$ is also the angle between R2 and the deviated ray (blue line) in the 1st prism, as illustrated in Fig. \ref{thetaH}:
\begin{equation}
\begin{aligned}
\theta_{H}=\frac{\alpha}{2}+\frac{\pi}{2}-\theta_{B}-\delta\theta&.  \label{thetaHeq1}
\end{aligned}
\end{equation}
Substituting Eq. \ref{apex} into Eq. \ref{thetaHeq1} obtains
\begin{equation}
\begin{aligned}
\theta_{H}=\pi-2\theta_{B}-\delta\theta=\alpha-\delta\theta,  \label{thetaHeq2}
\end{aligned}
\end{equation}
hence $DC=H\cos(\alpha-\delta\theta)=H\cos\alpha\cos\delta\theta+H\sin2\theta_{B}\sin\delta\theta$.\\

By substituting the above relations into Eq. \ref{EC}, the total optical path through the prism-pair is
\begin{equation}
\begin{aligned}
\boxed{P=(R+H\cos\alpha)\cos\delta\theta+H\sin2\theta_{B}\sin\delta\theta}
\label{finalpath1}
\end{aligned}
\end{equation}
and the optical phase experienced by frequency $\omega$ is
\begin{equation}
\begin{aligned}
\boxed{\varphi=\frac{\omega}{c}P}.
\label{phase}
\end{aligned}
\end{equation}

These expressions for the optical path $P$ and optical phase $\varphi$ are exact and can be easily used for numerical calculation of the spectral phase $\varphi(\omega)$ and its frequency derivatives GD=$d\varphi/d\omega$, GDD=$d^2\varphi/d\omega^2$ and higher order dispersion terms. Furthermore, since the prism-pair has two independent degrees of freedom $R,H$, it may allow to tune independently two orders of dispersion. For example, by searching for a specific $R,H$ configuration (and proper selection of the prism material), one may be able to compensate for both GDD and third order dispersion, which is important for management and generation of extremely short (sub 50fs) pulses. The refraction angle $\delta\theta$ for each wavelength/frequency can be calculated according to Snell's law by using the prism's apex angle and the Sellmeier formula for $n(\lambda)$ or $n(\omega)$.

\section{Approximation of the prism angular dispersion} \label{angulardispersion}

Although Eq. \ref{finalpath1} and Eq. \ref{phase} already allow complete calculation of the dispersion properties, much intuition can be gained by further developing the expressions with the assumption that the angle $\delta\theta(\lambda)$ is small. Replacing $\cos\delta\theta\approx1-(\delta\theta^{2}/2)$ and $\sin\delta\theta\approx\delta\theta$ yields
\begin{equation}
\begin{aligned}
\boxed{P\approx(R+H\cos\alpha)\left(1-\frac{\delta\theta^{2}}{2}\right)+(H\sin2\theta_{B})\delta\theta}.
\label{finalpath2}
\end{aligned}
\end{equation}
Note that Eq. \ref{finalpath1} is a generalization of the method presented in \cite{prismpairapprox}, where only the special case of $H=0$ was presented.

\begin{figure}
\centerline{\includegraphics[width=8cm]{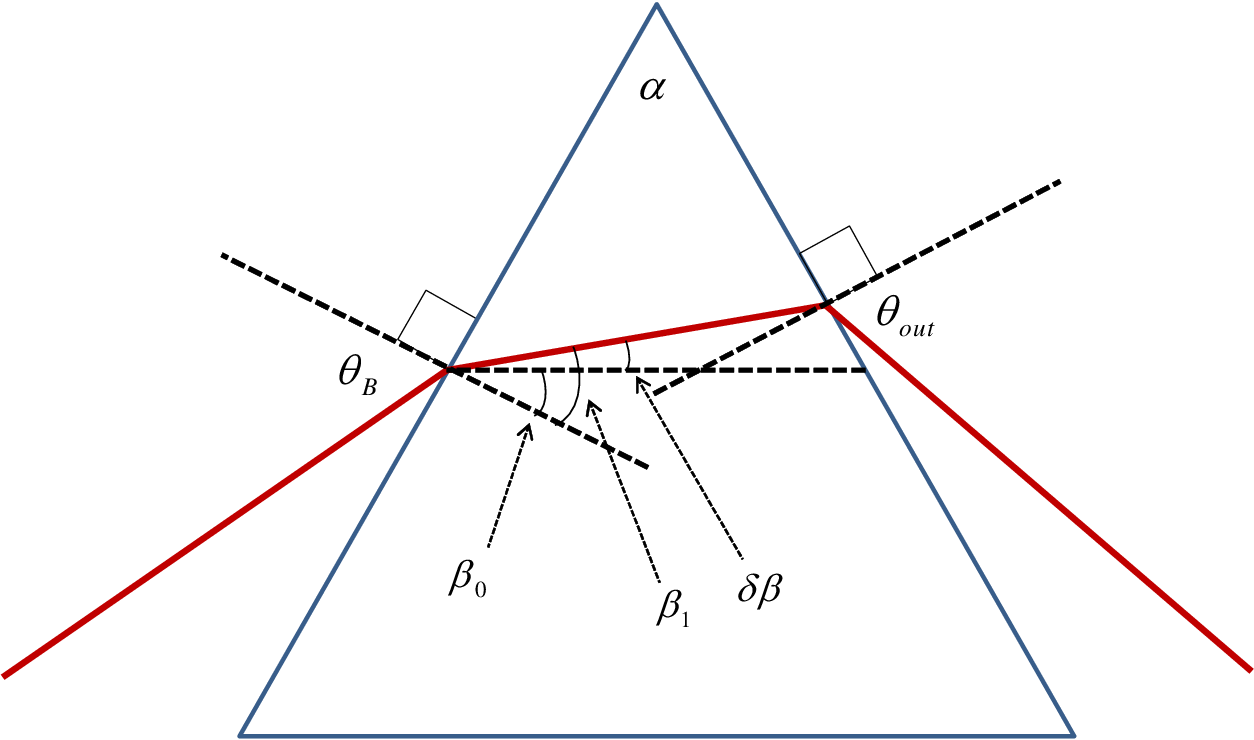}}
\caption{\label{smallangles} Definition of the refraction angles for small angle approximation of the angular dispersion.}
\end{figure}

The angle $\delta\theta$ can now be expressed using Snell's law assuming small angles, such that: $\cos\delta\theta\approx1$ and $\sin\delta\theta\approx\delta\theta$. It is assumed that the beam enters the 1st prism at the minimum deviation angle for a certain wavelength $\lambda_{0}$; that the entrance angle matches the Brewster angle $\theta_{B}$ for $\lambda_{0}$; and that the prism refractive index for $\lambda_{0}$ is $n(\lambda_{0})=n_{0}$. The refractive index for the deviated beam ($\lambda\neq\lambda_{0}$) is: $n=n_{0}+\delta n(\lambda)$. The angle of refraction inside the prism for the Brewster beam is $\beta_{0}$ (equal at both faces of the prism). As illustrated in Fig. \ref{smallangles}, the angle of refraction for the deviated beam inside the prism is $\beta_{1}=\beta_{0}+\delta\beta$. It is easy to show that the deviated beam will hit the exit face of the prism at an angle of $\beta_{2}=\beta_{0}-\delta\beta$. The exit angle for the deviated beam will be $\theta_{out}=\theta_{B}+\delta\theta$. In addition, since at Brewster angle $\tan\theta_{B}=n_{0}$ and $\theta_{B}=\pi/2-\beta_{0}$, the following relations hold: $\sin\theta_{B}=\cos\beta_{0}$ and $\cos\theta_{B}=\sin\beta_{0}$.\\

Following Snell's law from the entrance face to the exit, an approximation of the deviation angle $\delta\theta$ can be obtained. At the entrance:
\begin{equation}
\begin{aligned}
\sin\theta_{B}=n\sin\beta_{1}  \label{snell1}.
\end{aligned}
\end{equation}
Expanding the right hand side yields
\begin{equation}
\begin{aligned}
nsin\beta_{1}&=(n_{0}+\delta n)\sin(\beta_{0}+\delta\beta)\\
&=(n_{0}+\delta n)(\sin\beta_{0}\cos\delta\beta+\cos\beta_{0}\sin\delta\beta)\\
&\approx(n_{0}+\delta n)(\sin\beta_{0}+\delta\beta\cos\beta_{0})\\
&=n_{0}\sin\beta_{0}+\delta n\sin\beta_{0}+\delta\beta n_{0}\cos\beta_{0}\\
&=n_{0}\cos\theta_{B}+\delta n\cos\theta_{B}+\delta\beta n_{0}\sin\theta_{B}.
\label{snell1detail}
\end{aligned}
\end{equation}
Substituting into Eq. \ref{snell1} and dividing by $\sin\theta_{B}$ (and noting that $\tan\theta_{B}=n_{0}$) provides
\begin{equation}
\begin{aligned}
1+\frac{\delta n}{n_{0}}+n_{0}\delta\beta=1\\
\boxed{n_{0}^{2}\delta\beta=-\delta n}  \label{firstrelation}.
\end{aligned}
\end{equation}
Following similar logic at the exit face of the prism, obtains
\begin{equation}
\begin{aligned}
n\sin\beta_{2}=\sin\theta_{out}.
  \label{snell2}
\end{aligned}
\end{equation}
Expanding both sides of Eq. \ref{snell2} yields
\begin{equation}
\begin{aligned}
n\sin\beta_{2}&=(n_{0}+\delta n)\sin(\beta_{0}-\delta\beta)\\
&\approx n_{0}\sin\beta_{0}+\delta n\sin\beta_{0}-\delta\beta n_{0}\cos\beta_{0}\\
&=n_{0}\cos\theta_{B}+\delta n\cos\theta_{B}-\delta\beta n_{0}\sin\theta_{B}\\
\sin\theta_{out}&=\sin(\theta_{B}+\delta\theta)=\sin\theta_{B}\cos\delta\theta+\cos\theta_{B}\sin\delta\theta\\
&\approx\sin\theta_{B}+\delta\theta\cos\theta_{B}.
\label{snell2detail1}
\end{aligned}
\end{equation}
Equating both sides yields
\begin{equation}
\begin{aligned}
n_{0}\cos\theta_{B}+\delta n\cos\theta_{B}-\delta\beta n_{0}\sin\theta_{B}=\sin\theta_{B}+\cos\theta_{B}\delta\theta
\label{snell2detail2}
\end{aligned}
\end{equation}
Dividing Eq. \ref{snell2detail2} by $\sin\theta_{B}$ provides
\begin{equation}
\begin{aligned}
1+\frac{\delta n}{n_{0}}-n_{0}\delta\beta=1+\frac{\delta\theta}{n_{0}}\\
\delta n-n_{0}^{2}\delta\beta=\delta\theta \qquad. \label{secondrelation}
\end{aligned}
\end{equation}
Substituting Eq. \ref{firstrelation} into Eq. \ref{secondrelation} yields
\begin{equation}
\begin{aligned}
\boxed{\delta\theta=2\delta n=2(n-n_{0})}.  \label{finalrelation}
\end{aligned}
\end{equation}
Finally, Eq. \ref{finalrelation} can be substituted into Eq. \ref{finalpath1} and Eq. \ref{phase} to retrieve the total wavelength dependent phase through the prism-pair. Since the wavelength/frequency dependence is now only in $\delta n$, a simple expression for the GDD=$d^2\varphi/d\omega^2$ can be derived:
\begin{equation}\label{GDDapprox}
\begin{aligned}
\frac{d^2\varphi}{d\omega^2}=\frac{2}{c}\left(B-2A\delta n\right)\left(2\frac{dn}{d\omega} + \omega\frac{d^2n}{d\omega^2}\right) - \frac{4\omega}{c}A\left(\frac{dn}{d\omega}\right)^2,
\end{aligned}
\end{equation}
where $A=R+H\cos\alpha$ and $B=H\sin\theta_B$. The wavelength dependent refractive index of the prisms $n$ can be obtained from the Sellmeier formula\footnote{\url{http://refractiveindex.info/?shelf=glass&book=BK7&page=SCHOTT}}. The exact spectral phase and any of its derivatives ($d^{m}\varphi/d\omega^{m}$) can be easily calculated using the chain rule \cite{Cojocaru:03}
\begin{equation}
\begin{aligned}
\label{chainRule}
\frac{dn}{d\omega}&=-\frac{\lambda^2}{2\pi c}\frac{dn}{d\lambda}\\
\frac{d^2n}{d\omega^2}&=\frac{\lambda^3}{2\pi^2c^2}\left(\frac{dn}{d\lambda}+\frac{\lambda}{2}\frac{d^2n}{d\lambda^2}\right).\\
\end{aligned}
\end{equation}

For typical optical bandwidths it is safe to assume that $B\gg 2A\delta n$ and that $R\gg H$ ($R\sim10-50cm$ and $H\sim10mm$ are common values). Thus, it is intuitive to think of $A$ as the distance between the prisms ($A\approx R$), and of $B$ as the propagation distance inside the prisms. We can substitute Eq.\ref{chainRule} and $\omega=2\pi c/\lambda$ into Eq.\ref{GDDapprox} and obtain after some algebraic manipulation:

\begin{equation}
\begin{aligned}
\frac{d^2\varphi}{d\omega^2}\approx B\frac{\lambda^3}{\pi c^2}\frac{d^2n}{d\lambda^2} - A\frac{2\lambda^3}{\pi c^2}\left(\frac{dn}{d\lambda}\right)^2
\label{phaseapprox}
\end{aligned}
\end{equation}

The geometric dispersion in the second part attributes always a negative GDD for any material (independent of the sign of $dn/d\lambda$), whereas the first part depends on the material dispersion, which may be either positive (as is usually the case in the visible or NIR range), or negative (as is the case for most materials in the IR). Hence, a simple prism-pair offers tunable GDD, both negative and positive, for visible and NIR wavelengths, but for the IR range it will commonly provide only negative GDD. Tuned positive dispersion in the IR range can still be obtained by inserting a $1\times1$ telescope between the prisms that can flip the sign of the geometric dispersion by imaging the first prism beyond the 2nd prism, effectively generating a ``negative distance'' $R$ between the prisms \cite{shaked2015octave}.

\section{Summary} \label{summary}

 The performance of the Brewster prism-pair was reviewed - a common major component of ultrafast spectroscopy apparati. The total spectral phase was calculated as accumulated by broadband light in passage thorough the prism-pair. Following a simple and intuitive approach using the concept of wavefronts, we showed how the total optical path through a prism pair can be reduced to the path of a non-continuous ray, hence avoiding unnecessary computations of multiple refractions inside the prisms. Finally, the total phase and any of its derivatives can be easily calculated. Specifically, by careful choice of the two degrees of freedom of the prism pair R and H, it is possible to compensate two orders of dispersion simultaneously (e.g. $d^{2}\varphi/d\omega^{2}$ and $d^{3}\varphi/d\omega^{3}$), which is important to maintain high temporal resolution when using ultra-broadband light \cite{1367-2630-17-7-073024}.


\end{document}